\documentclass[12pt,sort&compress]{article}

\usepackage[dvips]{color}
\usepackage{amsmath}
\usepackage{graphicx} 
\usepackage[dvips]{color}
\usepackage{graphicx}
\usepackage{tabularx}
\usepackage{subfig}
\usepackage{cite}
\usepackage{amsfonts}
\usepackage[outercaption]{sidecap}    
\usepackage{hyperref}

\begin{document} 
\title{Steady states in a non-conserving zero-range process with extensive rates as a model for the balance of  selection and mutation}
\date{}
\author{Pascal Grange\\
Department of Mathematical Sciences\\
 Xi'an Jiaotong-Liverpool University\\
111 Ren'ai Rd, 215123 Suzhou, China\\
\normalsize{{\ttfamily{pascal.grange@xjtlu.edu.cn}}}}
\maketitle\

\begin{abstract}
We consider a non-conserving zero-range process with hopping rate proportional to the number of particles at each site. 
 Particles are added to the system with a site-dependent creation rate, and removed from the system with a uniform annihilation rate.    
 On a fully-connected lattice with a large number of sites,  the mean-field geometry  leads to a negative binomial law for the  number of particles at each site, with parameters depending on the hopping, creation and annihilation rates. This model of particles is mapped to a model of population dynamics: the site label is interpreted as a level of fitness, the site-dependent creation rate is interpreted as a selection function, and the hopping process is interpreted as the introduction of  mutants. In the limit of large density, the fraction of  the total population occupying each site approaches the limiting distribution in the house-of-cards model of selection-mutation,
 introduced by Kingman. A single site can be occupied by a macroscopic fraction of the particles if the mutation rate is below a critical value (which matches the critical value worked out in the house-of-cards model). This feature generalises to classes of selection functions that increase sufficiently fast at high fitness. The process can be mapped to a model of evolving networks, inspired by the Bianconi--Barab\'asi model, but involving a large and fixed set of nodes. Each node forms links at a rate biased by its fitness, moreover links are destroyed at a uniform rate, and redirected at a certain rate. If this redirection rate matches the mutation rate, the number of links pointing to nodes of a given fitness level  is distributed as the numbers of particles in the non-conserving zero-range process.
  There is a finite critical redirection rate if the density of quenched fitnesses goes to zero sufficiently fast at high fitness.

\end{abstract}
\pagebreak
\tableofcontents

\section{Introduction and background}

 Condensation is a feature of  steady states of a variety of out-of-equilibrium systems, including 
 granular materials, traffic flows and distributions of wealth.
   Some models of non-equilibrium statistical mechanics  with particles distributed over a large number of sites
 can  exhibit condensation as the macroscopic occupation of a single site \cite{bialas1997,majumdarMass,ZRPReview}. 
  A prominent class of such models is based on  the zero-range process (ZRP) \cite{spitzer5interaction,jamming,EvansBraz,DrouffeSimple},
 in which particles hop from any site at a rate $u$ depending only on the number of particles $n$
 at this site. If  $u$ is an increasing function, no condensate can form. However,  rates of the form
 \begin{equation}\label{decreasing}
 u_{b}( n ) = \left( 1 + \frac{b}{n}\right) \theta(n),
 \end{equation}
 where $\theta$ is the Heaviside step function, lead to the formation of a condensate if $b>2$.
  Once it is formed, the condensate can undergo an ergodic motion, or be  trapped at a site in the case of inhomogeneous hopping
 rates \cite{godreche2003,condensationDynamics,grosskinsky2003condensation,condensationInhomogeneous,coarsening}.\\

 In \cite{angel2005critical,nonConserving}, a non-conserving version of the ZRP 
 was introduced, with particles added to each site at a 
 constant rate, and particles removed from each site at a rate increasing (as a power law) with the number 
 of particles present at the site. The functional form of these rates was chosen to be the same
 for all sites. The hopping rate was 
  of the decreasing  form given in Eq. \ref{decreasing}. The model was studied on a large lattice using  a mean-field approximation. The phase diagram, which includes a super-extensive high-density phase,   
 was  worked out in terms of the scaling behaviour of the hopping current at 
 large system size. \\

On the other hand, Kingman \cite{KingmanSimple}  introduced a deterministic measure-valued 
 model of  the competition
 between selection and mutation in the fitness distribution 
  in a large haploid population, which exhibits a condensation
 phenomenon if the mutation rate is small enough. 
  The fitness of individuals is modelled as a single bounded number:
  at generation $n$, the fitness  distribution  is a probability measure $p_n$
 on the interval $[0,1]$.  
 The model assumes that the next generation consists of a fraction $\beta$ of mutants,
 whose fitness distribution is a fixed probability measure $q$ on  $[0,1]$,
 and of  descendants of the previous generation, contributing 
a skewed term proportional to $(1-\beta)xp_n(dx)$. The factor of $x$ reflects the higher reproduction rate of individuals with 
 higher fitness. Normalisation of the measure $p_{n+1}$ 
   induces the recurrence relation 
 \begin{equation}\label{measureProcess}
 p_{n+1}(dx) = (1-\beta) \frac{xp_n(dx)}{ \int_{[0,1]} xp_n(dx)} + \beta q(dx).
 \end{equation}
 If the mutation rate $\beta$ is lower than a critical value (depending only on the 
 mutant fitness $q$), the limiting distribution  develops an atom at the maximum value of fitness:
 \begin{equation}\label{pInf}
 p_\infty(dx) =  \frac{\beta q(dx)}{1-x} + \left(1- \frac{\beta}{\beta_c}\right) \delta_1,
\;\;\;\;{\mathrm{if}}\;\;\beta <\left(  \int_{[0,1]}\frac{q(dx)}{1-x} \right)^{-1}.
 \end{equation}
This model is termed  the house-of-cards model, 
 as mutations  reshuffle the genomic deck: the steady distribution of fitness is a skewed version of the mutant fitness $q$.   
The emergence of the condensate exhibits 
 universality properties depending only on the local behaviour of the mutant fitness
  $q$ at high fitness \cite{DereichEmergence,shape2018}. 
  Making the house-of-cards model more realistic involves the introduction of randomness (see  \cite{Yuan2019} for rigorous developments on random mutation rates, and \cite{Lenski} for applications to the Lenski experiment
  studying the fitness of a growing bacterial population through regular sampling). The introduction of new mutants,
 as well as the births and deaths of individuals, can be modelled as Markovian processes, and the steady 
  state of the system  could be characterised by the probability law of the number of individuals at each fitness level.
 This would allow for instance to estimate the fluctuations of the population at each level of fitness.\\

 In this paper  
 we consider  a non-conserving zero-range process  
   with a large number of sites, and map it to a model of selection and mutation.
   Hopping rates model mutation,  inhomogeneous 
  creation rates model selection, and homogeneous annihiliation rates model death. The hopping rates
  are not assumed to be of the form given in Eq. \ref{decreasing}, but are chosen to be 
 proportional to the number of particles:
\begin{equation}\label{un}
 u(n) := \beta n,
\end{equation}
 where the factor $\beta$ models the mutation rate. 
   From a population-dynamics viewpoint, this choice corresponds to the assumption 
 that the rate of introduction of new mutants is proportional to the current population. 
  From a particle viewpoint, this choice corresponds to considering particles as independent random walkers 
 (when a hopping event happens, a particle is drawn uniformly from  the set of 
 particles in the entire system, hence the probability for a given fitness level to be the departure
 site is proportional to the number of particles present at this site).\\

    Such extensive hopping rates could not lead to a condensate in the case of the  conserving ZRP \cite{ZRPReview}.
  In the limit of a large number $L$ of sites, the site labels (divided by $L$) 
   can be thought of as fitness levels that can approach any value in the interval $[0,1]$. The introduction of randomness 
 in the evolution of the system 
 avoids the partition of the population into generations.  Moreover, it should
  allow to work out the probability law of the number of individuals at each fitness level in a steady state. 
 If mutants are allowed to hop from a site to any other site,
  each site has many neighbours in the large-$L$ limit, and approximations 
 inspired by mean-field theory are expected to give good results.
   Moreover, hopping events  in the ZRP can be interpreted in terms of redirection events of links 
   in a network \cite{angel2005critical,nonConserving}, using a mapping from links pointing to nodes to a system 
  of particles \cite{BB}. The present model of selection and mutation can therefore be mapped to a
 model of a network (with nodes endowed with quenched fitnesses).\\

%

 In Section 2 we will describe the model more completely 
  and set up notations. In Section 3 we will write the steady-state  master equation, 
 assuming  the numbers of particles at all sites to be independent. In Section 4 we will 
  solve this  equation and the steady-state  numbers of particles  will appear to be 
  negative binomial variables, with site-dependent parameters. 
 In Section 5 we will work out  the average fraction of all particles occupying 
  each site in a simple case, and identify a regime of parameters in which the average 
 density goes to infinity, while a finite fraction of the particles is concentrated at the highest fitness value.
 The ratio of the  average number of individuals at a given fitness level to the average density will be 
 related to the skewed large-time distribution appearing in the deterministic house-of-cards model (Eq. \ref{pInf}).
 In Section 6 we will use the mapping from networks to particles in the Bianconi--Barab\'asi model \cite{BB,BBCond} 
   to propose an analogue of our model in terms of directed links on a large network, which can be created, annihilated and 
 redirected. The rates of these processes will be adjusted  to make the analogy complete. 
  In Section 7 the model will be generalised, based on features depending only on the local behaviour (at high fitness) of the creation rate 
 and mutant fitness.

%
%
%
%
%

\section{Non-conserving ZRP with extensive, inhomogeneous rates}

Consider a lattice of $L$ sites, with site labelled $l$  carrying a random number $n_l$ of particles. 
  We use the site label to model the fitness level of individuals in a haploid population: there are 
  $n_l$ individuals with fitness level $l/L$ (so that in the large-$L$ limit the fitness can be arbitrarily close 
 to any value in the interval $[0,1]$).
 The number $n_l$ can evolve due to  annihiliation, creation  
 and hopping from site to site. These three processes model deaths, births and mutations in the population.
 The rates of the processes (i.e. their probabilities per unit of time) are chosen as follows.\\

 Particles are annihilated at each site at  a rate  proportional to the number of particles 
 at the site,  with a proportionality factor $\delta>0$,  independent 
 of the site label:
 \begin{equation}\label{deathRate}
 n_l  \rightarrow  n_l - 1\;\;{\mathrm{with\;rate}}\;\; \delta n_l,\;\;\;{\mathrm{for\;all}}\;\;n_l>0.
 \end{equation}
 Particles are created at site labelled $l$, at a rate chosen 
 to be an increasing positive function $f$ of  the fitness level $l/L$:
 \begin{equation}\label{creationRate}
 n_l  \rightarrow  n_l + 1\;\;{\mathrm{with\;rate}}\;\;  f( l/L)( n_l + 1),\;\;\;{\mathrm{for\;all}}\;\;n_l\geq 0.
 \end{equation}
  where the shift in the factor $(n_l+1)$  is introduced in order to prevent 
 the state with no particles at any site from being steady. One can think of this shift as modelling
 the action of an external agent, who introduces one particle at any empty site, at a site-dependent rate 
   adjusted to maintain the creation  rates of the selection process. One can also think that the creation 
   of a new particle  happens at site labelled $l$ at a cost that is inversely proportional to the 
 number of particles present at the site after the creation (and this cost is a decreasing function of the fitness of the site).\\

 The function $f$ will be referred to as 
 the selection function, as it models the higher reproduction rate of individuals with higher fitness.  
  The unit of the quantities $\beta$ , $\delta$ and $f$  (introduced in Eqs \ref{un}, \ref{deathRate}, \ref{creationRate} )
   is the inverse of a time, because numbers of particles 
  and their probabilities are dimensionless quantities. 
  We may choose a particular process and set its time scale as the unit of time 
 for the model.  Let us assume  that one particle  is created 
on average per unit of time at the site of maximum fitness, labelled $l=L$, if this site contains no particle.
 This choice of time scale is equivalent to the choice $f(1)=1$ 
for the maximum value of the selection function.\\

 The hopping process is a zero-range process: particles
 hop from site labelled $l$ at a rate depending only on the number of particles present at the
 site.  Let us choose an extensive hopping rate:
 \begin{equation}\label{hoppingRate}
 u( n ) = \beta n,\;\;\;\;\beta >0.
 \end{equation} 
  The positivity constraint on the number of particles that has to be imposed 
 through a factor of $\theta(n)$ in the decreasing hopping rate of Eq. \ref{decreasing}, is automatically 
 satisfied.\\

 When a particle hops from site $l$, the destination site  
 is chosen randomly among the other $L-1$ sites, with a probability 
 law derived from a fixed probability measure $(q_m)_{1\leq m \leq L}$, 
  i.e. $q_m\geq 0$ and $\sum_{m=1}^L q_m=1$.  
 In the large-$L$ limit we will pick these numbers
 as special values of a smooth probability density $q$ on the interval $[0,1]$:
 \begin{equation}\label{mutantDensity}
 q_l:= \frac{q\left(l/L \right)}{\sum_{j=1}^L q\left( j/L \right) }\simeq_{L\rightarrow\infty} 
\frac{1}{L} q\left( l/L\right).
 \end{equation}
  The corresponding hopping processes from site $l$ to site $m\neq l$ are therefore described by:
 \begin{equation}\label{hoppingProcess}
 (n_l, n_m) \rightarrow  (n_l - 1, n_m + 1)\;\;{\mathrm{with\;rate}}\;\; \beta n_l \frac{q_m}{ 1- q_l},
 \end{equation}
 where the denominator in the rate ensures normalisation of the probability law of the 
 destination site. These processes model the production of mutants in the population, 
  and the  density $q$ is the probability density of the fitness of the new mutants. It will be referred to as the mutant 
  density. Moreover, we will assume that $q(1)=0$ (new mutants have zero probability 
 of having maximum fitness).\\


 The list of parameters of the model therefore consists of a large integer $L$, two positive numbers 
  $\beta$ (the mutation rate) and $\delta$ (the death rate, or annihilation rate), 
 a smooth probability density $q$ (the mutant density) on the interval $[0,1]$, satisfying $q(1)=0$,
  and a positive 
increasing function $f$  (the selection function) on the interval $[0,1]$, 
 satisfying $f(1)=1$.

\section{Mean-field analysis of the model and steady-state equations}
Let us postulate that the steady-state probability of each configuration of the system factorises: we  
 assume the existence of $L$ probability distribution functions, denoted by $(p_l)_{1\leq l \leq L}$, 
 such that 
 \begin{equation}\label{facto}
P(  n_1,\dots, n_L) = \prod_{l=1}^L p_l( n_l).
 \end{equation} 
 The conserving zero-range process is known to exhibit such a factorised steady-state probability distribution \cite{EvansBraz}.
    In the present case 
  the destination site in the hopping process  (Eq. \ref{hoppingProcess}) is drawn from the 
 set of neighbours  of the departure site. This set  becomes infinite at large $L$ as the probability 
 density  $q$ is smooth. We are therefore in the mean-field geometry.  The source of randomness in the model 
 is the same as in the dynamics of  urn models 
  \cite{godreche2003,urn1,urn2}.  The 
 factorisation property of Eq. \ref{facto} will therefore hold as a result of the large-$L$ limit.\\


 For a fixed site labelled $l$,  let us write schematically the steady-sate master equation as
\begin{equation}\label{masterEquation}
\frac{dp_l}{dt}(n)= 0 = A_l(n) + C_l(n)+M_l(n),\;\;\;\;\forall n \geq 0, 
\end{equation}
where the  annihilation, creation and mutation terms are denoted 
 by  $A_l$, $C_l$ and $M_l$ respectively.\\

 Based on the translation-invariant 
  death rates of Eq. \ref{deathRate}, the annihilation term reads
\begin{equation}
\begin{split}
A_l(n) &= \delta \left\{  (n+1)  p_l(n+1) -   n \theta(n) p_l(n) \right\},\\
\end{split}
\end{equation}
where the factor $\theta(n)$ imposes that there 
  should be at least one particle on site before annihilation  takes place (even though this constraint is redundant
 because of the factor of $n$ from the annihilation rate).\\

Based on the fitness-dependent creation rates of Eq. \ref{creationRate}, the creation term reads 
\begin{equation}
C_l (n)=  f( l/L) \left\{\theta(n)  n p_l(n-1) -  (n + 1 ) p_l(n) \right\},
\end{equation}
where the factor $\theta(n)$ imposes that there should be 
  at least one particle on site after creation has taken place.\\

In the large-$L$ limit, the flow of particles to site $l$ per unit of time 
 is proportional to $q(l/L)$ and to the average density $\rho$
 of the system:
\begin{equation}\label{density}
\begin{split}
 \rho := \frac{1}{L}\sum_{k=1}^L \overline{n_k},\;\;\;\;\;\; \overline{n_k}:= \sum_{n\geq 0 } n p_k( n).
\end{split}
\end{equation}
 Indeed the  contribution of the normalisation factors  in the hopping rates to site labelled $l$ 
   (Eqs \ref{mutantDensity} and \ref{hoppingProcess}) are negligible in the large-$L$ limit:
 \begin{equation}\label{meanFieldFlux}
\begin{split}
 \beta q_l\left( \sum_{k\neq l } \sum_{n_k\geq 0} \frac{1}{1- q_k} n_k p_k(n_k) \right)& \simeq_{L\rightarrow \infty}
  \beta\frac{1}{L} q(l/L)\sum_{k\neq l } \left( \overline{n_k} +  \frac{1}{L} \overline{n_k}q\left( k/L\right) + o(L^{-1}) \right)\\
 &= \beta \rho q\left( l/L\right) + o(1).
\end{split}
 \end{equation}
  The mutation term in the steady-state master equation therefore reads:
\begin{equation}
\begin{split}
M_l(n)=   \beta \left\{ ( n+1) p_l( n+1 ) - n \theta( n ) p_l( n ) \right) + \rho q\left( l/L\right)\left(  \theta( n ) p_l( n-1) -  p_l( n) \right\},\\
\end{split}
\end{equation}
 where the factors of $\theta( n )$   impose there should be least one particle 
 on site when a particle hops from the site (even though this constraint is redundant 
 in one case because of the factor of $n_l$ in the hopping rate).\\

There are therefore four terms with constraints and four terms without constraint in each of the steady-state conditions:
 \begin{equation}\label{balanceEq}
\begin{split}
 0 = \;& \theta( n ) \left\{  \left( \beta \rho q\left( l/L \right) +   f\left(l/L \right) n \right)p_l(n-1) 
  - ( \delta  +  \beta ) n p_l(n)    \right\}\\
   &+   (  \beta+ \delta)(n+1)p_l(n+1)  -  \left( \beta \rho q\left( l/L\right) +   f\left( l/L \right)( n+1)\right)p_l( n), \;\;\;\forall n \geq 0.
 \end{split}
 \end{equation} 
 For $n=0$   the steady-state condition  reduces to
 \begin{equation}
 p_l(1) =   \frac{ \beta \rho q\left( l/L\right) +  f(l/L)}{ \beta+\delta} p_l(0),
 \end{equation}
 so that at  any value of $n$ the constrained and unconstrained 
 parts of the balance equation (Eq. \ref{balanceEq}) are separately equal to zero.
 This reproduces the structure of the mean-field master equation derived in \cite{nonConserving}
 in the case of homogeneous rates, where only one probability law needs to be determined
 to express the probability of any configuration. 
  By  induction on  $n$ we therefore obtain:
\begin{equation}
 p_l(n ) =  \prod_{k=1}^n\frac{  \beta \rho q\left( l/L\right) +  f(l/L)k   }{ (\beta + \delta) k}  p_l(0).
\end{equation} 
 The normalisation factors $\left(p_l(0)\right)_{1\leq l \leq L}$ and the 
 average density $\rho$ still  have to be fixed. 

\section{Normalisation and average density}
Let us factorise the selection function in the expression of $p_l(n)$ and 
  introduce the Pochhammer symbol
\begin{equation}
(a)_n := a\times(a+1)\times\dots\times(a+n-1) = \frac{\Gamma( a+n)}{\Gamma( a )}.
\end{equation}
\begin{equation}
p_l(n) = p_l(0) \prod_{k=1}^n \frac{\beta\rho q( l/L) +  f(l/L)k}{(\delta +\beta)k} = p_l(0) \left(\frac{f(l/L)}{\delta + \beta}\right)^n\frac{1}{n!}
 \left( \frac{\beta\rho q( l/L)}{f(l/L)} + 1 \right)_n.
\end{equation}
 
Using the identity 
\begin{equation}\label{negativeBinomialParams}
\frac{1}{(1-z)^a} = 1 + \sum_{n\geq 1} \frac{(a)_n}{n!} z^n,\;\;\;\;|z|<1,
\end{equation}
we can express the normalisation factor at site labelled $l$,
 provided $ \delta + \beta >1$ (which condition ensures convergence of the sum at all sites because  
  $f(1)=1$ is the maximum of the selection function).  Defining a parameter $\zeta>0$ by
\begin{equation}\label{zetaDef}
 \delta + \beta = 1+\zeta,
 \end{equation}
  we obtain the normalisation factor (in terms of the still-unknown  density $\rho$) as:
\begin{equation}\label{predictedDensities}
 p_l(0) = \left( 1 - \frac{f(l/L)}{1+\zeta}\right)^{\frac{\beta\rho q( l/L)}{f(l/L)} + 1}.
\end{equation}

Moreover, we can recognise $p_l$ as a negative binomial law
 \begin{equation}
p_l(n) =   \frac{1}{n!}( r_l)_n  \pi_l^{r_l} ( 1 -\pi_l )^n,
\end{equation}
 with  parameters $\pi_l$ and $r_l$ depending on the fitness level:
\begin{equation}\label{paramDef}
\pi_l= 1 - \frac{f(l/L)}{1+\zeta},\;\;\;\;\;\; r_l= 1 + \beta\rho \frac{q( l/L)}{f( l/L)}.
\end{equation}
 We deduce an expression of the mean value of the number of particles at site labelled $l$,
 in which the only unknown parameter is the average density:
\begin{equation}
 \overline{n_l}=\sum_{n>0} n p_l(n)=   r_l\frac{1-\pi_l}{\pi_l},
\end{equation}

\begin{equation}\label{nBarl}
  \overline{n_l}=   \beta\rho \frac{q(l/L)}{1+\zeta - f( l/L)} +\frac{f( l/L)}{1+\zeta - f( l/L) }.
\end{equation}
 We can already see that for $l<L$, the first term (which is 
 a skewed version of the mutant density) will be dominant if the density $\rho$ is large.
 Moreover, the assumption $q(1) = 0$ implies that the number of particles at the 
 maximum fitness level follows a geometric law, and $\overline{n_L} = \zeta^{-1}$.
    Values of $\zeta$ larger than $1$
  can therefore be considered large for our purposes, and we observe that values of $\zeta$ close to zero
  yield  large numbers of particles at maximum fitness. Indeed the site labelled $l=L$ does not receive any particle from 
 the mutation process, and at this site $\beta+\delta = 1+\zeta$ combine as an effective total annihilation 
  rate, while the local creation rate is $f(1)=1$.\\

 Consistency with the definition of the average density $\rho$ (Eq. \ref{density}) yields,
 rewriting  Riemann sums as integrals using the large-$L$ limit:
\begin{equation}\label{rhoFixing}
\begin{split}
 \rho &= \frac{1}{L}\sum_{l=1}^L \frac{f\left(l/L \right)}{1+\zeta} \left( \beta\rho 
 \frac{q\left(l/L \right)}{f\left(l/L\right)}+1 \right)\left( 1 - \frac{f(l/L)}{1+\zeta}\right)^{-1}\\
&\simeq_{L\to\infty}   \beta\rho \int_0^1 \frac{q(x) dx}{1+\zeta - f(x)} + \int_0^1\frac{f(x)dx}{1+\zeta - f(x)}.
\end{split}
\end{equation}
 As $\zeta>0$ and $f(1)=1$ is the maximum of the selection function, all the integrands in the above expression are positive. 
 The average density $\rho$ can therefore only be positive if 
\begin{equation}
 \beta \int_0^1 \frac{q(x) dx}{1+\zeta - f(x)} < 1.
 \end{equation}
  Considering the parameter  $\zeta$  as fixed, we can rewrite this condition (using  Eq. \ref{zetaDef}) as a lower bound on the death rate:
\begin{equation}\label{deltaCrit}
\delta > \delta_c(\zeta),\;\;\;\;{\mathrm{with}}\;\;\;\delta_c(\zeta) = 1 + \zeta - \left( \int_0^1 \frac{q(x)dx}{1+\zeta - f(x)}\right)^{-1}.
\end{equation}
 The average density can therefore be expressed in terms of
 the mutant density, selection function, and  two parameters 
  $\zeta$ and $\epsilon$ that depend only on the pair $(\beta,\delta)$:
\begin{equation}\label{rhoExpr}
\rho =\frac{1}{\epsilon}\left( \int_0^1 \frac{q(x)dx}{1+\zeta - f(x)}\right)^{-1}    \int_0^1\frac{f(x) dx }{1+\zeta - f(x)},\;\;\;\;\;\; 
\zeta = \delta +\beta -1 >0, \;\;\;\;\; \epsilon = \delta - \delta_c( \zeta)>0.
\end{equation} 
 
\section{Example: linear selection function}

In the house-of-cards model of selection and mutation \cite{KingmanSimple}, the individuals that do not undergo mutation
 have a number of descendants that is proportional to their fitness. This  motivates us to choose
 \begin{equation}
 f(x) : = x.
\end{equation}
We would like to define the mutant density $q$ so that the integral in the definition of   $\delta_c(\zeta)$ in Eq. \ref{deltaCrit}
 has a finite limit when $\zeta$ goes to zero. Otherwise $\delta_c(0)$ would equal $1$ and the corresponding
 mutation rate would be zero. For this purpose it is enough to choose $q$ with the following local
 behaviour at high values of fitness:
\begin{equation}\label{localq}
q( 1-h) = O( h^\alpha), \;\;\;\;{\mathrm{where}}\;\;\;\alpha > 0.
\end{equation}
With such a choice of mutant density, $\beta$ goes to a strictly positive limit  if the parameter $\zeta$ goes to zero at fixed $\epsilon>0$,
  hence the notation:
\begin{equation}\label{betaCrit}
 \beta_c := \left( \int_0^1  \frac{q(x)dx}{1-f(x)}\right)^{-1},\;\;\;\;\;\lim_{\zeta\to 0,{\mathrm{fixed}}\;\epsilon} \beta = \beta_c - \epsilon.
\end{equation}
  We recognise $\beta_c$ as the critical value of the mutation 
 rate that appears  in the house-of-cards model \cite{KingmanSimple} (see Eq. \ref{pInf}).\\

 The average  density $\rho$  can be expressed for this particular choice of $f$ using Eq. \ref{rhoExpr}. 
  It diverges logarithmically when the parameter $\zeta$ goes to zero:
\begin{equation}
\rho ( \epsilon, \zeta) =\frac{1}{\epsilon}\left( \int_0^1 \frac{q(x)dx}{1+\zeta - x}\right)^{-1} \left(    -(1+\zeta) \log\left( \frac{\zeta}{1+\zeta} \right)  -1      \right),
\end{equation} 
 \begin{equation}\label{largeRho}
 \rho( \epsilon, \zeta)  \sim_{\zeta\rightarrow 0^+} -\frac{1}{\epsilon}\beta_c\log\zeta.
\end{equation}
  Consider the average number of particles at site labelled $l = xL$, for $x$ in $[0,1]$, denoted in the large-$L$ limit by $\bar{n}(x)$, 
 divided by the average density. Its expression consists of two terms.  
  One is absolutely continuous with respect to the mutant density $q$, even at $\zeta = 0$,  and the other
  converges to a Dirac
 measure at maximal fitness value when $\zeta$ goes to zero:
\begin{equation}\label{measureProc}
 \frac{\overline{n}( x )}{\rho( \epsilon, \zeta) } = \left( \left( \int_0^1  \frac{q(x)dx}{1+\zeta-x} \right)^{-1}  -\epsilon\right) \frac{q(x)}{ 1 +\zeta - x} + \frac{1}{\rho( \epsilon, \zeta) } \frac{x}{1+\zeta -x}.
\end{equation}
Indeed,  if $\phi$ is a smooth test function on the interval $[0,1]$, integrating by parts yields:
 \begin{equation}\label{testIntro}
\begin{split}
-\frac{1}{\log \zeta} \int_0^1 \frac{x\phi( x )}{1+\zeta - x}dx & =  -\frac{1+\zeta}{\log \zeta} \left( -\int_0^\frac{1}{1+\zeta} \phi\left( (1+\zeta) x \right) dx + \int_0^{\frac{1}{1+\zeta}}\frac{\phi\left((1+\zeta) y \right)}{1-y} dy\right)\\
& =- \frac{1}{\log \zeta} \left( -\log\left( 1 - \frac{1}{1+\zeta}\right) \phi( 1 )\right) +o(1)\\
& = \phi( 1 ) + o(1), \;\;\;\;(\zeta \to 0 ).
\end{split}
 \end{equation}

Taking the limit  of the expression \ref{measureProc} when $\zeta$ goes to zero (at fixed $\epsilon$, using 
Eqs \ref{betaCrit},\ref{largeRho}) yields\\
\begin{equation}\label{densities}
 \lim_{\zeta \to 0,{\mathrm{fixed}}\;\epsilon}
\frac{\overline{n}( x )}{\rho(\epsilon, \zeta) } = \left( \beta_c-\epsilon\right) \frac{q(x)}{1- x} + \frac{\epsilon}{\beta_c} \delta_1(x),
\end{equation}
 where we recognise the steady-state measure $p_\infty$ in the house-of-cards model (Eq. \ref{pInf}), because 
 the quantity $\beta_c-\epsilon$ is the limit of $\beta$ when $\zeta$ goes to zero at fixed $\epsilon$ (see Eq. \ref{betaCrit}).\\


  The variance of the negative binomial distribution yields the following expression for the
 variance of the number of particles at site labelled $l$ (using the expression of Eq. \ref{paramDef} for the parameter $\pi_l$):
\begin{equation}\label{varianceEq}
 {\mathrm{Var}}(n_l)=   \frac{{\overline{n_l}}}{\pi_l} =  \frac{(1+\zeta) \overline{n_l}}{1+\zeta - f( l/L) }.
\end{equation}
This expression yields the scaling of fluctuations of the population at fitness level $l=xL$, for small 
 values of the parameter $\zeta$:
 \begin{equation}\label{fluctuationBounds}
 \frac{\Delta n(x)}{\overline{n(x)}}=\sqrt{\frac{(1+\zeta)}{(1+\zeta - x)\overline{n(x)}}}
\sim_{\zeta \to 0}\sqrt{\frac{-\epsilon}{\beta_c( \beta_c-\epsilon) q(x) \log\zeta}}.
\end{equation}
 This scaling is a manifestation of the law of large numbers.
 At low values of $\zeta$  the average density $\overline{n_l}$  is large (from Eq. \ref{densities}, we see
 that it is proportional to the large total density $\rho(\epsilon,\zeta)$).
 The random population at fitness level labelled $l$ is therefore in the situation of the law of 
 large numbers: there are a large number of independent random walkers in the 
 system, and each of them carries  a Boolean random variable which equals one if and only if  the walker is 
 at site labelled $l$. The random variable $n_l$ is the sum of these random Booleans. Its average 
  $\overline{n_l}$ is proportional to the total number of walkers. 
  For independent random walkers, the variances of the random Booleans add up to the 
 variance of $n_l$. At large density this induces the
 scaling $\sqrt{{\mathrm{Var}}(n_l)}/\overline{n_l} = O(\overline{n_l}^{-1/2})$, which goes to zero 
 as $(-\log\zeta)^{-1/2}$, as read off from Eq. \ref{fluctuationBounds}.\\
 
 Because of the local behaviour of the mutant density close the maximum fitness value (Eq. \ref{localq}),
 these fluctuations diverge at fixed death and mutation rates when $x$ goes to 1. The fluctuations 
 are therefore concentrated around the highest fitness value.

\section{Mapping to a model of a large network with quenched fitnesses}

 Condensation phenomena involving the same type of integrals as in the 
   criterion of Eq. \ref{pInf}  have arisen in network science. This is no accident, 
  as models of growing networks such as 
 the {\hbox{Bianconi--Barab\'asi}} model of competition for formation of links  \cite{BB}
    can be mapped to models of particles  \cite{BBCond}. Consider a network, with nodes that can connect to each other 
 by forming directed links. 
   Each node in the network is endowed with  a certain level of fitness representing the aptitude of the node to attract new links.
 The network is mapped to a system of particles as follows. 
 Each node is mapped to a state, endowed with the same level of fitness. The degeneracy
  of the fitness levels leads to a density of states.
   The states are populated with particles: a  particle is added to each state for each 
 link pointing to the corresponding node in the network. If  the network is grown by adding new nodes, 
    each forming a given number of links, preferential attachment to connected  links of higher fitness
  can be  understood \cite{BBCond} in terms of  condensation of particles at the level of highest 
  fitness.\\

   This mapping from networks to particles  puts our model in perspective
 with evolving networks.  Instead of adding new nodes one by one and letting them form 
  links to existing nodes, let us consider a fixed, large system of nodes.  Our network can therefore evolve
    only by changing the configuration of links.
    More precisely, consider again $L$ regularly spaced fitness levels  in the interval $[0,1]$.
   Let us initialise our network by introducing $v_l$ nodes with fitness level $l/L$  (for all $l$ in $[1..L]$), with 
 \begin{equation}\label{vl}
 v_l = [ q_l V] + 1,
 \end{equation}
  where the square bracket denotes the integral part, and $V$ 
 is a large integer. The shift of 1 is added so that there is exactly one
  node at maximum fitness, as we imposed  $q(1)=0$.
   If  $L$ is large and $V$ is much larger than $L$,  Eq. \ref{vl} implies that the total number of  nodes in the network is close to $V$.
    Moreover  the fraction of nodes with 
   fitness $l/L$  is close to $q_l$.
    This large, fixed system of nodes  corresponds to 
  a situation with a large number of states,  and a  density of states given by $q$.\\

 
 Let us label  each of the $v_l$  nodes at fitness level $l/L$ by an integer $s$ in $[1..v_l]$,
 and let us denote by $\lambda(l,s)$ the number of links that point to  node labelled $s$.  
  Let us  map 
 oriented links to particles as in the Bianconi--Barab\'asi model. We denote by $m_l$ the number of particles at fitness level $l/L$ (eventually this quantity will 
  be identical to $n_l$, but let us keep different notations until we have described the dynamics completely).
  The number $m_l$ is the  sum of the numbers of links pointing to the $v_l$ states at fitness level $l/L$:
\begin{equation}\label{populationMapping}
  m_l = \sum_{s=1}^{v_l}\lambda( l, s).
\end{equation}
  Let us keep the set of nodes (and their fitnesses) fixed. We have a large network with a density of quenched fitnesses given by $q$.
     A configuration of this network   is   given by the collection $(\lambda( l, s)_{1\leq s \leq v_l})_{1\leq l \leq L}$.
  Such a configuration is mapped to a configuration of populations of particles grouped by fitness levels, $(m_l)_{1\leq l \leq L}$,
 through Eq. \ref{populationMapping}.\\

  Let us allow the configuration of links in the network to change through three elementary processes:\\
- creation of a link (this process creates a particle at the state corresponding to the node to which the created link is pointing);\\
- destruction of a link (this process destroys a particle at the state corresponding to the node to which the  destroyed link was pointing);\\
- redirection of an existing link to a different node (this process makes a particle hop from a state to another state, and preserves the total number of links and particles, see \cite{angel2005critical} for a model of a network with redirection events induced by a ZRP with a decreasing hopping rate).\\

 We can adjust the rates of these processes so that 
 they induce a dynamics for the family of random numbers of particles $(m_l)_{1\leq l \leq L}$  
  that reproduces the dynamics  of  the populations denoted by $(n_l)_{1\leq l \leq L}$ in the 
   non-conserving ZRP.\\

  The easiest rates to adjust are the destruction rates. Let us assume that each link in the network is destroyed
  at the rate $\delta$:
 \begin{equation}
 \lambda( l, s) \rightarrow \lambda( l, s) - 1  \;\;{\mathrm{with\;rate}}\;\; \delta,\;\;\;{\mathrm{for\;all}}\;\;\lambda( l, s)>0,\;\; l \in [1..L],\;\;s \in [1.. v_l].
\end{equation}
  Summing these rates over $s$ at fixed $l$  yields a destruction rate  $\delta m_l$ for the particles at fitness level $l$.\\

  Let us assume that each link in the network is redirected at the rate $\beta$,
  and that 
  the new node it points to is drawn uniformly from the set of possible nodes.
   Consider two distinct fitness values $l/L$ and $k/L$. The redirection events inducing the hopping event  
  $(m_l,m_k) \rightarrow (m_l-1, m_k+1)$ are the events in which one of the $m_l=\sum_{s=1}^{v_l} \lambda(l, s)$ links
  pointing to one of the $v_l$  nodes with fitness $l/L$ is redirected to one of the $v_k$ nodes of fitness $k/L$.
  As the new destination node is drawn uniformly when a redirection event happens, the rates of these events sum to
 \begin{equation}
     \beta m_l\frac{v_k}{ V - v_l} \underset{V\to\infty}{\simeq}  \beta m_l\frac{q_k}{ 1 - q_l}, 
 \end{equation}
 where  we have used Eq. \ref{vl} for the expression of $v_l$ 
 in terms of the mutant density. This reproduces the rates of the hopping processes described in  Eq. \ref{hoppingProcess}.
  Neglecting $q_l$ in the denominator by taking the large-$L$ limit of these rates in the r.h.s. amounts to neglecting redirection events between nodes 
 of identical fitness (which do not induce hopping of particles between different fitness levels). Redirection events generically induce hopping
 of particles in the mean-field geometry.\\

 For the creation rates to match those of Eq. \ref{creationRate},  consider the creation 
 process of one particle at the maximum fitness level (which contains exactly one node or state because $v_L = [q(1) V] + 1 = 1$),
   when this level contains no particle. Let us set the  rate of this process to one,
  to define the time scale of the evolution of the network (this is the fastest creation process in the network when it is in the configuration
  without any link).
  Then, biasing the creation of new links by an increasing function $f$ with $f(1) = 1$,
  let us pick the rates
  \begin{equation}
 \lambda( l, s) \rightarrow \lambda( l, s) + 1  \;\;{\mathrm{with\;rate}}\;\;   f( l/L) \left( \lambda(l,s) + \frac{1}{v_l}\right),\;\;{\mathrm{for\;all}}\;\;\lambda( l, s)\geq 0,\;\; l \in [1..L],\;\;s \in [1.. v_l].
\end{equation}
  Each of the above rates is the sum of the rates of  $(V-1)$ different processes (each corresponding to a different choice 
 for the origin of the new link). We assumed that all these processes have the same rate (the origin of the new link is drawn uniformly). 
 Hence we can write the 
 above expression in terms of quantities that depend only on the node to which the new link is pointing.
 The term  $\frac{1}{v_l}$ in the r.h.s.  implies that the destination site of the new link 
  is chosen uniformly among the $v_l$ states with fitness $l/L$ when the first link to this group of nodes is created.
   Summing these rates over $s$ at fixed $l$  yields a creation rate  $f( l/L) \left(  m_l + 1\right)$ for the 
  particles of fitness $l/L$.\\

 This model of an evolving network with a large and fixed set of nodes (and a density of  quenched fitnesses
  equal to the mutant density $q$)
   is  therefore mapped to our non-conserving ZRP. The network
 evolves through uniform death rate  $\delta$ of links, uniform redirection rate  $\beta$ of links, and
 creation rate biased by $f$  towards nodes of higher fitness.
    With these rates, the number of links to nodes with fitness $l/L$ satisfies the same 
 master equation as the population $n_l$. We can therefore conclude that it follows a negative binomial 
  law at steady state. Moreover, the total density of links in the  network diverges logarithmically when the parameter 
 $\zeta = \delta + \beta -1$ goes to zero. It should be noted that the quantity denoted by $\beta$  is quite distinct from the 
 thermodynamic quantity $\beta_{th}$ that can be introduced to map the fitness $x$ to a Boltzmann factor for an energy $E$ through
  the change of variables $x=e^{-\beta_{th} E}$. Such a change of variables is known to induce a Bose occupation factor based on the 
  factor $(1-x)^{-1}$ in the expressions of  Eq. \ref{pInf}.  Bose--Einstein condensation can occur at equilibrium  in quantum systems 
 of particles, at high $\beta_{th}$ (low temperature) for a
  certain  fixed 
 density of states  in the space of energies.  In our model the density of states is fixed in the space of fitnesses, and larger
 values of $\beta$ correspond to larger rates of agitation (in the form of rewiring of the network), which work against the emergence
 of a condensate. 
  Condensation at highest fitness occurs in the limit of a large density 
 if the redirection rate $\beta$ is below a critical value that depends only on the distribution $q$ of quenched fitnesses.\\

\section{Generalisation}

 Going back to the expression of the average density $\rho$ of the system (Eq. \ref{rhoExpr})
 for a more general choice of selection function $f$, we observe that an atom developed at the highest fitness value 
 in the case of a linear selection function because the integral 
 \begin{equation}
  J(f,\zeta) = \int_0^1  \frac{f(x) dx}{1+\zeta - f(x)},
 \end{equation}
 which is present in the numerator in the expression of the average density (Eq. \ref{rhoExpr}), 
 goes to infinity when $\zeta$ goes to zero. This divergence is entirely  due to the local behaviour 
 of $f$  at high fitness values.\\ 

 We can therefore generalise the above results to a one-parameter family of 
 selection functions:
 \begin{equation}
  f_\chi(x) = 1 - (1- x)^\chi,\;\;\;{\mathrm{for}}\;\; \chi>0,
 \end{equation}
 or to any selection function with the same local behaviour around $1$.
 The mutant density must satisfy 
$q( 1-h) = O( h^\alpha)$ with  $\alpha > \chi -1$,
  so that the critical value $\beta_c$ of the mutation rate  is  finite.\\

The dominated convergence theorem implies that the difference
\begin{equation}\label{singularity}
  J(f_\chi,\zeta) - \int_0^1  \frac{ dh}{\zeta + h^\chi} = - \int_0^1  \frac{ h^\chi dh}{\zeta +  h^\chi}
 \end{equation}
 has a finite limit when $\zeta$ goes to zero. The case $\chi=1$   studied in Section 5 gives rise to a logarithmic 
 divergence, and for $\chi<1$ the dominated convergence theorem implies that the average density (Eq. \ref{rhoExpr}) 
has a finite limit when $\zeta$  goes to $0$ at fixed $\epsilon$.\\
 
 For $\chi>1,$  the rate of divergence of the quantity  $J(f_\chi,\zeta)$ for small $\zeta$ is 
  the same as that of the subtracted integral in the l.h.s. of Eq. \ref{singularity}. This integral  diverges as $\zeta^{\frac{1}{\chi}-1}$
 when $\zeta$ goes to zero, as can be seen by factorising $\zeta$ in denominator and changing variables:
 \begin{equation}
  \int_0^1  \frac{ dh}{\zeta + h^\chi} = \frac{1}{\zeta}\int_0^1  \frac{ dh}{ 1+ \left( \frac{h}{\zeta^{\frac{1}{\chi}}} \right)^\chi} = 
\zeta^{\frac{1}{\chi}-1}\int_0^{\zeta^{-\frac{1}{\chi}}} \frac{du}{1 + u^\chi} \sim_{\zeta \to 0} \zeta^{\frac{1}{\chi}-1} \int_0^{+\infty} \frac{du}{1 + u^\chi}.
 \end{equation}
  The  integral in the above expression can be worked out  in terms of the beta function $B$ 
  using the change of variable defined by $1 + u^\chi = y^{-1}$.
 \begin{equation}
\begin{split}
 \int_0^{+\infty} \frac{du}{1 + u^\chi} &= \frac{1}{\chi}\int_0^1 t^{-\frac{1}{\chi}}( 1 -t)^{\frac{1}{\chi}-1}dt\\
  &= \frac{1}{\chi}B\left( 1-\frac{1}{\chi}, \frac{1}{\chi}\right) = \frac{1}{\chi}\Gamma\left( 1-\frac{1}{\chi}\right)\Gamma\left(\frac{1}{\chi}\right) = 
 \frac{\pi}{\chi \sin{\frac{\pi} {\chi}}},
\end{split}
 \end{equation}
 where we used the Euler  reflection formula, $\Gamma( x ) \Gamma( 1-x) = \frac{\pi}{\sin(\pi x)}$.\\

%
%
 At a fixed value of the parameter $\epsilon$,  Eq. \ref{rhoExpr} therefore implies that for a selection function equal to $f_\chi$ with $\chi >1$, the density goes to infinity 
 as a power law when $\zeta$ goes to zero:
\begin{equation}\label{rhoGenEst}
 \rho(\epsilon,\zeta) \sim_{\zeta\to 0}\frac{\beta_c}{\epsilon} \frac{\pi}{\chi \sin{\frac{\pi} {\chi}}} \zeta^{\frac{1}{\chi} - 1}. 
 \end{equation}

 To generalise the emergence of an atom at maximum fitness, let us introduce a smooth test function $\phi$ (as in Eq. \ref{testIntro}).
   The difference 
\begin{equation}
\int_0^1\frac{\phi(x) f_\chi(x)dx}{1+ \zeta - f( x) } 
-\int_0^1\frac{\phi(x) dx}{1+ \zeta - f_\chi( x) } = \int_0^1 \frac{\phi(1-h) h^\chi}{\zeta  + h^\chi  } dh 
\end{equation} 
 has a finite limit when $\zeta$ goes to zero. Hence an equivalent of  the l.h.s.  when $\zeta$ goes to zero 
 is given by the second term.  Let us  work out an equivalent of this term by the changes of variables defined by $x=1-h$, followed by $h = 1- \zeta^{\frac{1}{\chi}} u$, 
  and a Taylor expansion of the test function around 1:
\begin{equation}
\begin{split}
 \int_0^1\frac{\phi(1-h)dh}{\zeta + h^\chi } &=  \zeta^{\frac{1}{\chi} - 1} \int_0^{\zeta^{-\frac{1}{\chi}}} \frac{\phi(1- \zeta^{\frac{1}{\chi}}u )du}{1 + u^\chi} = 
  \zeta^{\frac{1}{\chi} - 1} \int_0^{\zeta^{-\frac{1}{\chi}}} \frac{( \phi(1 ) + o(1))du}{1 + u^\chi} \\
 & \sim_{\zeta\to 0} \phi(1)\zeta^{\frac{1}{\chi} - 1}  \int_0^{+\infty} \frac{du}{1 + u^\chi }=   \frac{\pi}{\chi \sin{\frac{\pi} {\chi}}} \zeta^{\frac{1}{\chi} - 1} \phi(1).
\end{split}
\end{equation} 
 Hence we find a generalisation
 of Eq. \ref{densities} for  the ratio of the average number of  particles at fitness $x$ to the average density:
\begin{equation}
\begin{split}
  &\lim_{\zeta \to 0,{\mathrm{fixed}}\;\epsilon}
\frac{\overline{n}( x )}{\rho(\epsilon, \zeta) } = (\beta_c-\epsilon)\frac{q(x)}{1- f(x)} + \frac{\epsilon}{\beta_c} \delta_1(x),\\
(1-f( x ) )&\sim_{x\to 1}  (1-x)^\chi,\;\;\; \chi >1,\;\;\;\;q(1-h) = O(h^\alpha),\;\;\;\;\alpha > \chi - 1.
\end{split}
\end{equation}


 The fluctuations can again be expressed using Eq. \ref{varianceEq}, and the 
 asymptotic expression for the density (Eq. \ref{rhoGenEst}), for $f=f_\chi$, with $\chi > 1$:
 \begin{equation}
 \frac{\Delta n(x)}{\overline{n(x)}}
\sim_{\zeta \to 0}\sqrt{\frac{\epsilon\chi \sin{\frac{\pi} {\chi}}}{\pi\beta_c( \beta_c-\epsilon)q(x)}} \zeta^{\frac{\chi - 1 }{2\chi}},
\end{equation}
 which is again consistent with the law of large numbers, which applies to the number 
 of particles at a given level of fitness when the average density of the system is large.

\section{Summary and discussion}

 We have studied the steady states of  a non-conserving zero-range process with extensive 
 hopping, creation and annihilation rates, on a fully-connected lattice with a large number of sites.
    This model can  be interpreted naturally as a stochastic model of the balance between selection and mutation
  in a haploid population. 
   Site labels model the bounded fitness. 
  Site-dependent creation rates model selection, and 
  the hopping process with extensive rates models the introduction of new mutants.
   Assuming that the probability of each configuration factorises (which is asymptotically valid for a large number 
 of sites as the model is in the mean-field geometry), we established that  the number of particles at each site is distributed according 
 to a negative binomial law, with site-dependent parameters.  The 
  average  density of the system can be expressed in the steady state in terms of integrals
 of the mutant density and selection function. The average population at each fitness level in $[0,1[$ 
 is dominated at large density by a skewed version of the mutant density.\\

 In the limit of large density, the relative fluctuation of the population at each fitness
 level in $[0,1[$ goes to zero. The limit of large density
 is controlled by  the parameter we denoted by $\zeta$, which must be positive for a steady state
 to be reached, and equals $\beta + \delta - 1$, where $\beta$ is the mutation rate and $\delta$ is 
 the annihilation rate. 
  The inverse of $\zeta$ equals the average number 
 of individual  at the maximum fitness level.
   The rate of divergence of the average density of the system  (when the parameter $\zeta$ goes to zero)
 has been found to depend only 
  on the local behaviour of the selection function at maximal fitness.\\
 
 In the  region of the $(\beta,\delta)$ plane close to the half-line of equation 
$\beta + \delta = 1$, $\beta < \beta_c$, a non-conserving ZRP with extensive
 hopping rate exhibits a macroscopically large number of particles in the 
  level of highest fitness.
 This is in contrast with the conserving ZRP, where condensation is known to occur only 
 for decreasing hopping functions. However, the birth process is biased towards higher
   fitness through the selection function, and its rates also grow macroscopically.
  Moreover, for  a  fixed destination  site labelled $m$, the rate of the hopping process
 from a departure site $l\neq m$ (defined in Eq. \ref{hoppingProcess}), divided by the population $n_l$ at the departure site,
 equals $q_m(1-q_l)^{-1}$, which is minimised if the departure site is labelled $l=L$.
   Moreover, the expression of the critical value 
 $\beta_c$ is identical to the one worked out in the house-of-cards model. The critical value is therefore strictly positive 
  if the mutant density goes to zero sufficiently fast at high fitness.\\





 Considering the entire population with random evolution processes (with ancestors 
  contributing to the distribution of fitness until they die), is more realistic than 
  the approach of the house-of-cards model  in which each individual is assigned a generation label. Moreover,
 the present approach allows for an explicit estimate of the density of the population
 even if the case of non-linear selection function $f$. It has been appreciated 
 that the local behaviour of $q$ at maximum fitness value is responsible for the
 emergence of condensation \cite{DereichEmergence}, but the derivations rely strongly 
 on the distribution of the moments of the distributions $p_n$ and $q$ at all orders.
 These moments  emerge naturally from the normalisation of  the measure process (Eq. \ref{measureProcess}).
 The mean-field approach makes use of the thermodynamic limit in two ways: 
 through the large number of sites $L$, which allows to take the continuum 
 limit of the values of fitness, and through the large average number of particles in the system,
 that controls the fluctuations of the population at each level of fitness.\\


  Moreover, mapping from networks to systems 
   of particles as in the Bianconi--Barab\'asi model allows to interpret 
  the population at a given fitness level as the number of links to nodes of the same fitness
   in a large  network, with a fixed set of nodes. This model of a network has one more 
 parameter than the non-conserving ZRP, which is the total number of nodes. This number is assumed 
   to be sufficiently large   for the ditribution 
 of  quenched fitnesses to approach the mutant density.
   The selection function is mapped to a preferential attachment rate to connected nodes of higher fitness.
    Links are allowed to be destroyed at a uniform rate, and to be
   redirected (which processes are absent from the Bianconi--Barab\'asi model, in which the network is grown by adding 
 nodes and links). The mutation rate is mapped to the redirection rate, which completes the analogy. 
  A higher redirection rate  can be thought of as  a higher agitation rate. 
 If the density of quenched fitnesses goes to zero sufficiently fast at high fitness.
 there is a finite critical value of the redirection rate, below which a finite fraction 
 of a large steady population of links condensates at the node of highest fitness.

\section*{Acknowledgements}

It is a pleasure to thank Linglong Yuan for numerous discussions.



\bibliography{bibRefs} 
\bibliographystyle{ieeetr}
\end{document}